\documentstyle[12pt,aaspp4]{article} 
\def\tempest%
{\begin{array}{ccc} 
1 & 1 & 1 \\ 
1 & 1 & 1 \\ 
4 & 3 & 8 
\end{array}} 

\def\kpc{{\rm kpc}}

\begin{document}

\title{The Ratio of Total to Selective Extinction Toward Baade's Window}
\author
{Andrew Gould, Amelia Stutz, and Jay A.\ Frogel}
\affil{Ohio State University, Department of Astronomy, 
140 W.\ 18th Ave., Columbus, OH 43210, USA}
\authoremail{gould@astronomy.ohio-state.edu,stutz@astronomy.ohio-state.edu,
frogel@astronomy.ohio-state.edu}
\begin{abstract} 

	We measure the ratio of total to selective extinction,
$R_{VI}\equiv A_V/E(V-I)$, toward Baade's Window by comparing the $VIK$
colors of 132 Baade's Window G and K giants from Tiede, Frogel, \& Terndrup
with the solar-neighborhood $(V-I),(V-K)$ relation from Bessell \& Brett.  
We find $R_{VI}=2.283 \pm 0.016$, and show that our measurement has no 
significant dependence on stellar type from G0 to K4.
Adjusting the Paczy\'nski et al.\
determination of the centroid of the dereddened Baade's Window clump for 
this revised value of $R_{VI}$, we find $I_{0,\rm RC}=14.43$ and 
$(V-I)_{0,\rm RC}=1.058$.  This implies a distance to the Baade's Window
clump of $d_{BW} = 8.63\pm 0.16\,\kpc$, where the error bar takes account
of statistical but not systematic uncertainties.


\end{abstract} 

\section{Introduction} 

	Paczy\'nski \& Stanek (1998) found that the $VIK$ colors of
clump giants in Baade's Window were anomalous in the sense that at
fixed $(V-K)_0$, they were redder in $(V-I)_0$ than clump giants in
the solar neighborhood by 0.2 mag.  Stutz, Popowski, \& Gould (1999) found
a similar offset (0.17 mag) for RR Lyrae stars in Baade's Window
compared to those in the solar neighborhood.  Popowski (2000) showed that 
part of these offsets was simply due to errors in the original photometry
used by both groups.  When he incorporated the revised OGLE photometry of
Paczy\'nski et al.\ (1999), he found that the offset in $(V-I)$ 
shrank to $\sim 0.11$ mag in both cases.  

	Popowski (2000) then reviewed the various attempts
to explain such an offset in terms of a difference between the intrinsic
properties of stars in the two populations, an idea advanced by 
Paczy\'nski (1998) and by Stutz et al.\ (1999).  He argued that such an
explanation was not impossible, but unlikely, and that a more
plausible explanation is that the ratio of total to selective extinction
$R_{VI}\equiv A_V/E(V-I)=2.5$ adopted by both Paczy\'nski \& Stanek (1998)
and Stutz et al.\ (1999) from Stanek (1996) was incorrect.  Both color
anomalies could be solved, he noted, by adopting $R_{VI}= 2.1$.
Popowski (2000) used this re-evaluation to draw various conclusions about
the extragalactic distance scale.

	In the course of calibrating a $K$-band clump-giant distance 
indicator and applying it to measure the distance modulus of the Baade's Window
bulge field, Alves (2000) measured
$R_{VI}=2.26$.  He did so by comparing the $VIK$ colors of his
20-star sample of Baade's Window clump giants taken from 
Tiede, Frogel, \& Terndrup (1995) with the $VIK$ colors of local clump
stars.  

	Here we improve on the Alves (2000) measurement by incorporating
a factor $\sim 7$ more stars into the analysis, i.e.,
all 138 G0-K4 giants with $VIK$ photometry from Tiede et al.\ 
(1995).  We determine $R_{VI}$ by comparing these $VIK$ colors 
to the $VIK$ colors of nearby stars as determined by Bessell \& Brett (1988).
We find,
\begin{equation}
R_{VI}\equiv {A_V\over E(V-I)}=2.283 \pm 0.016.
\label{eqn:rvi}
\end{equation}
This reduces the color anomalies to $\sim 0.05\,$mag and so
qualitatively confirms Popowski's (2000) explanation of them.

\section{Measurement of $R_{VI}$}

	To measure $R_{VI}$, we assume that the $VIK$ colors of giant stars in
Baade's Window as measured by Tiede et al.\ (1995) are intrinsically the
same as those in solar neighborhood as determined by Bessell \& Brett (1988).
(We
will partially test this assumption below.)\ \  Bessel \& Brett (1988) 
give $VIK$ colors from G0 to M5, that is, over the color range
$1.75\leq(V-K)_0 \leq 5.96$.  However, there is considerable evidence that
the spectral energy distributions of M giants (at fixed spectral type) 
differ significantly between Baade's Window and the solar neighborhood
(Frogel \& Whitford 1982, 1987; Tiede et al.\ 1995).  This does not
necessarily mean that the $(V-I),(V-K)$ color-color relation is different, 
and in fact we will
present evidence below that it is not.  However, to be conservative,
we restrict consideration to $(V-K)_0\leq 3.5$, which eliminates all M giants
and K5 giants as well.  That is,
\begin{equation}
1.75\leq(V-K)_0 \leq 3.50.
\label{eqn:besrange}
\end{equation}
We begin with the sample of 509 stars with
optical and infrared photometry from the BW4b field of Tiede et al.\ (1995).
Note that the columns headed ``$V_0$'' and ``$(V-I)_0$'' 
in that paper actually
give $V$ and $(V-I)$.  We recover the original $K$ magnitudes by adding
$K=K_{0,\rm Tiede}+0.14$, as indicated by Table 1 of Tiede et al.\ (1995).  
We then obtain $(V-K)_0$ using visual extinctions, $A_V$, from Stanek's (1996)
extinction map together with the relation,
\begin{equation}
A_K = 0.11\,A_V.
\label{eqn:avk}
\end{equation}
Note that the Stanek (1996) map (available by anonymous ftp at 
astro.princeton.edu, stanek/Extinction) has been corrected to the
zero point found by Gould, Popowski, \& Terndrup (1998) and
Alcock et al.\ (1998).  These authors made their zero-point determinations
by comparing the $(V-K)$ colors of local K giants and RR Lyrae stars, 
respectively, with the $(V-K)$ colors of similar stars 
in Baade's Window, making use of the extinction ratio given by equation
(\ref{eqn:avk}).  Hence, for consistency, we use the same ratio here.

	Of the original 509 Tiede et al.\ (1995) stars, 
185 lack $VI$ photometry.
Of the remainder, 86 lack $A_V$ measurements because they fall within $2'$
of NGC 6522 where Stanek (1996) found the $A_V$ determinations to be
unreliable.  A further 96 stars are bluer than the color interval
(\ref{eqn:besrange}), and an additional 4 stars are redder.  
This leaves a total of 138 stars.  

	For each of these 138 stars, we use linear interpolation to
estimate the $(V-I)_{0,\rm Bessel}$ predicted from
their measured $(V-K)_0$ and the Bessell \& Brett (1988) $VIK$ 
color-color relation.  We then fit the data to
a two-parameter model
\begin{equation}
(V-I)_{\rm Tiede} - (V-I)_{0,\rm Bessell}= \alpha A_V + \beta [(V-K)_0 -2.348],
\label{eqn:linfit}
\end{equation}
where the offset 2.348 is chosen to eliminate the correlation between 
$\alpha$ and $\beta$.

	We remove outliers as follows.  We do the fit using all the
data, and determine the ``errors'' by forcing $\chi^2$ per degree of freedom
to unity.  We find the largest $\sigma$ outlier, eliminate it, and repeat
the process.  We stop when the largest outlier is less than $3\,\sigma$.
This eliminates six outliers.  We find,
\begin{equation}
\alpha = 0.4380\pm 0.0031,\qquad \beta = -0.0023\pm 0.0127
\qquad [1.75\leq (V-K)_0\leq 3.50].
\label{eqn:alphabeta}
\end{equation}
Since $R_{VI}=\alpha^{-1}$, we obtain equation (\ref{eqn:rvi}). From the
fact that $\beta$ is consistent with zero at the $1\,\sigma$ level, we conclude
that the shift from $(V-I)$ to $(V-I)_0$ depends only on $A_V$ and not
on the color of the star.  We therefore set $\beta=0$ and show our resulting
fit in Figure \ref{fig:one}.  Note that since $R_{VI}$ will in general be a 
function of color or spectral type, our measurement should be taken as
applying to stars at the mean of our sample, $(V-K)_0=2.35$, i.e., K0 giants.
Our value, $R_{VI}=2.28$,
 is in good agreement with the one found by Alves (2000),
$R_{VI}=2.26$, and the one adopted by Tiede et al.\ (1995), $R_{VI}=2.25$.

	Also plotted on Figure \ref{fig:one} are the four stars ({\it open
circles}) that were excluded from the fit because they were too red.  Note
that they lie very close to the local $VIK$ curve of Bessell \& Brett (1988).
If we repeat the entire procedure including these four stars, we obtain
\begin{equation}
\alpha = 0.4384\pm 0.0033,\qquad \beta = 0.0073\pm 0.0092
\qquad [1.75\leq (V-K)_0\leq 5.96].
\label{eqn:alphabeta2}
\end{equation}
with four outliers excluded.
That is $R_{VI}=2.281\pm 0.017$, essentially identical to equation
(\ref{eqn:rvi}).  In either case, the slope $\beta$ is consistent with zero,
indicating that our determination is not a signficant function of spectral
type.

	This non-dependence on $(V-K)_0$ color appears to be in strong
conflict with Figure 14 of Tiede et al.\ (1995) which shows a large number
of points in the range $4<(V-K)_0<6$ that lie $\sim 0.3\,$ mag below the
Bessell \& Brett (1988) relation.  In fact, all the points with CCD
$VI$ photometry
from Tiede et al.\ (1995) lie close to the line (as they do in our Fig.\
\ref{fig:one}).  The remaining points are from Frogel, Whitford \& Rich 
(1984) and Frogel \& Whitford (1987) who obtained their own
single-channel infrared data, but relied on the earlier photographic data
of Whitford \& Blanco (1979) and Arp (1965) for the $VI$ photometry.
 From the small overlap between these older photographic data 
and the CCD photometry reported
in Tiede et al.\ (1995), we estimate that the earlier photographic photometry
may have a zero-point error of $\sim 0.3\,$ mag in $(V-I)$
in the sense of being too blue.  If this zero-point error is
confirmed by future observations, it would mean that the $(V-I),(V-K)$ 
color-color relation is the same for Baade's Window and the solar neighborhood,
but not necessarily the correspondance between colors anad spectral type.
Nevertheless, to be conservative, we base our results only on the G0--K4
sample.

\section{Discussion}

	As with essentially all methods for determining total and selective
extinction, our measurement relies on the assumption that the colors of
stars in Baade's Window are the same as those in the solar neighborhood.
If the mean $(V-I)_0$ color at fixed $(V-K)_0$ differed between the two
populations by $\Delta(V-I)$, then our estimate of $\alpha$ would likewise
be in error by $\Delta \alpha = \Delta(V-I)/\langle A_V\rangle$ where 
$\langle A_V\rangle=1.496$ is the mean value of $A_V$ over our final sample
of 132 stars.  However, from the fact that $\beta$ is consistent with zero
(eq.\ [\ref{eqn:alphabeta}]), such an offset would have to be independent of
spectal type from G0 to K4 (and arguably to M5).  This seems quite implausible.
In addition, approximately the same offset would have to apply to RR Lyrae
stars (Stutz et al.\ 1999).  Most probably, the fault lies not in the stars,
but in the dust.  

	Popowski (2000) reexamined the  Wo\'zniak \& Stanek (1996)
method by which Stanek (1996) determined $R_{VI}=2.5$.  Making use of
the original tests done by Wo\'zniak \& Stanek (1996), he found that this
determination depends in part on an initial assumption about $R_{VI}$ so
that values as low as $R_{VI}\sim 2.3$ would be consistent with the data.

	Our measurement of $R_{VI}$ removes most, but not all, of the anomalous
color problems found by Paczy\'nski \& Stanek (1998) and by Stutz et al.\ 
(1999).  In the latter case, Popowski's (2000) revised offset
$0.11\pm 0.02$ mag is now reduced to $0.04\pm 0.02$, and so is only a
$2\,\sigma$ discrepancy.  However, for the clump giant anomaly found by
Paczy\'nski \& Stanek (1998), the formal uncertainty is only 
$\sim 0.003$ mag, so statistical fluctuations do not provide a
plausible explanation.  Nevertheless, the remaining discrepancy is small
and may be due to a combination of small offsets between the 
photometric zero points of the various measurement systems that are used
to make the comparison.  

	Paczy\'nski et al.\ (1999) used $R_{VI}^{\rm Stanek}=2.50$ from
Stanek (1996) to deredden the
observed centroid of the clump in their $VI$ color-magnitude diagram of Baade's
Window.  The zero point of their $V$-band extinctions is based on Gould et al.\
(1998) and Alcock et al.\ (1998) and is not affected by the present paper.
However, the $I$-band photometry should be adjusted fainter by
$\Delta I_0 = (1/R_{VI}^{\rm Stanek} - 1/R_{VI}^{\rm GSF})
\langle A_V\rangle = 0.056$, where $R_{VI}^{\rm GSF}=2.283$ is the value we
determine here and $\langle A_V\rangle=1.48$ is the mean extinction of
clump stars measured by Paczy\'nski et al.\ (1999).  This adjustment yields
a clump centroid of $I_{0,\rm RC}=14.43$ and $(V-I)_{0,\rm RC}=1.058$.
The color is only 0.05 mag redward of the centroid of the Hipparcos clump
(Paczy\'nski et al.\ 1999).  Inserting the magnitude into Udalski's 
(2000) $I$-band calibration of the clump standard candle, 
$M_I= (-0.26 \pm 0.02) + (0.13 \pm 0.07)(\rm [Fe/H] + 0.25)$,
we obtain a distance
modulus to Baade's Window
\begin{equation}
\mu_{BW} = 14.43 - M_I = 14.68\pm 0.04,
\label{eqn:mubw}
\end{equation}
where we have adopted [Fe/H]$_{BW}=-0.15\pm 0.10$.  The error bar takes account
of all statistical uncertainties, 0.025 mag for $M_I$ (Udalski 2000),
$0.02$ mag for the observed brightness of the Baade's Window clump
(Paczy\'nski et al.\ 1999), $0.04\times (1-R_{VI})=0.022$ mag for the
zero-point uncertainty of the Stanek (1996) map (Alcock et al.\ 1998),
and $0.0033\times \langle A_V\rangle=0.005$ mag for the uncertainty in 
$R_{VI}$ (this paper), but does not take account of systematic errors.
Equation (\ref{eqn:mubw}) is in good agreement with
Alves' (2000) determination using the $K$-band clump distance indicator, 
$\mu_{BW}=14.58\pm 0.11$.

{\bf Acknowledgements}: 
We thank Piotr Popowski for his useful comments on the manuscript and 
David Alves for pointing out the error in the column headings of Tiede et al.\ 
(1995).  This work was supported in part by grant AST 97-27520 from the NSF.


\begin{figure}
\caption[junk]{\label{fig:one}
Color-color $(V-K)_0$ vs.\ $(V-I)_0$ plot of 142 Baade's Window
giant stars ({\it points}) together with
the color-color relation ({\it solid curve}) of solar-neighborhood stars 
taken from Bessell \& Brett (1988).  The $(V-K)_0$ color is determined
from $VK$ photometry of Tiede et al.\ (1995) corrected for extinction
according to Stanek (1996), assuming $E(V-K)=0.89\,A_V$. The $(V-I)_0$
color is determined in this paper by fitting for $R_{VI}=A_V/E(V-I)$,
essentially by moving the points vertically until they straddle the curve.
Four red stars ({\it open circles}) were excluded from the fit because of
concerns that the spectral energy distributions of Baade's Window M giants
are different from those of M giants in the solar neighborhood.
Six outliers ({\it crosses}) were also excluded from the fit.  
The arrow shows the effect
of 1 magnitude of $V$ band extinction.  Since this arrow is parallel to
the color-color curve, even large errors in the extinction cannot affect
the results.  
}
\end{figure}

\clearpage

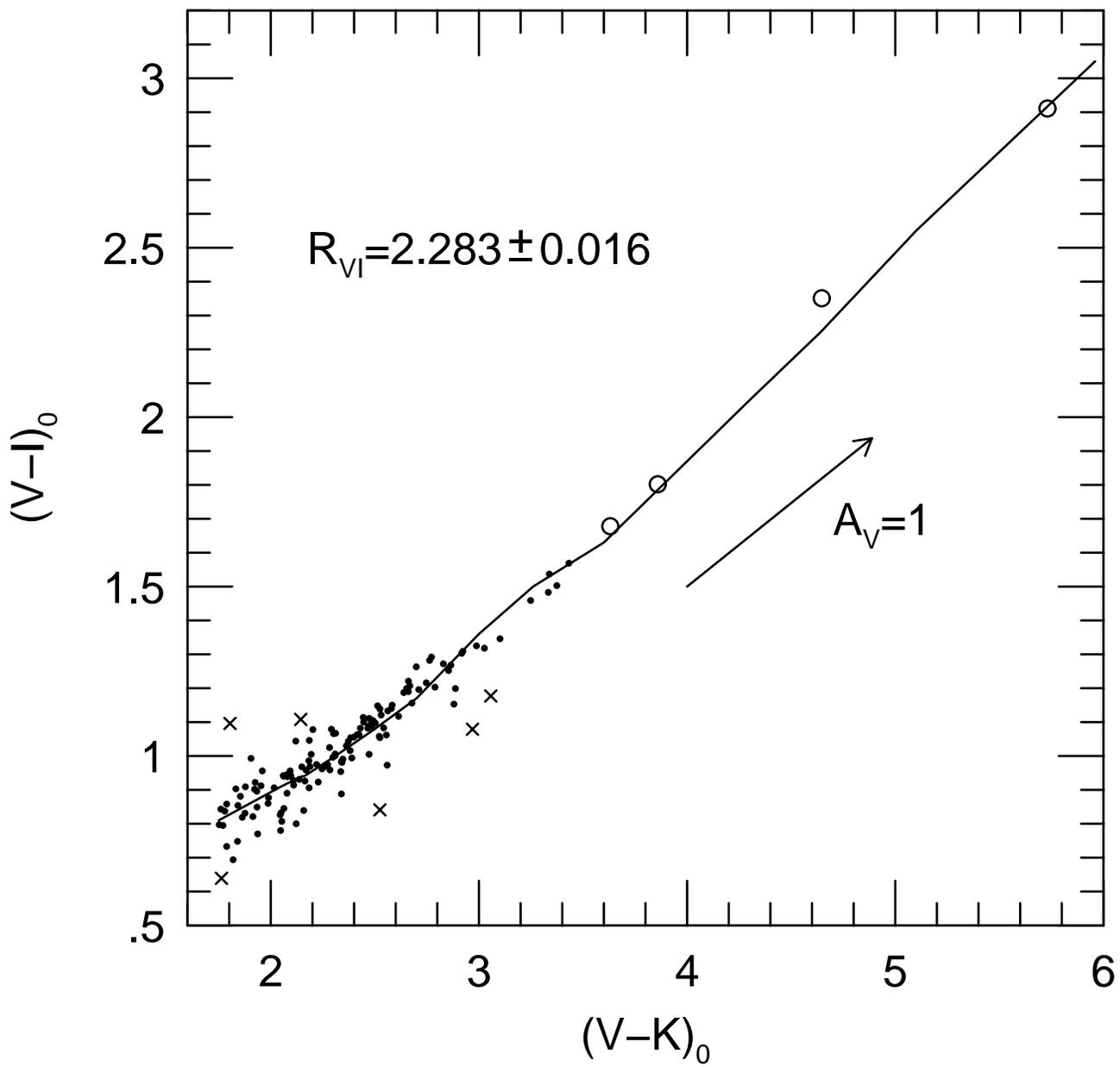
\end{document}